\documentclass[10pt,conference,letterpaper]{IEEEtran}
\usepackage{times,amsmath,epsfig}
\usepackage{subfig}
\usepackage{cite}
\usepackage{url}
\usepackage{float}

\usepackage{wrapfig}

\usepackage{array}
\title{Urban Mobility Scaling:\\Lessons from `Little Data'}
\author{%
{Galen Wilkerson{\small$^{1}$}, Ramin Khalili{\small$^{2}$}, Stefan Schmid{\small$^{2}$}}%
\vspace{1.6mm}\\
\fontsize{10}{10}\selectfont\itshape
$^{1}$ Technical University Berlin, Berlin, Germany\\
\texttt{galen.wilkerson@inet.tu-berlin.de}\\
$^{2}$ Technical University Berlin \& T-Labs, Berlin, Germany\\
\texttt{\{ramin,stefan\}@net.t-labs.tu-berlin.de}\\
\fontsize{9}{9}\selectfont\ttfamily\upshape
}
\begin{document}
\maketitle

\vspace*{-.4cm}

%WE MUST PURCHASE DATA !

%CONSISTENT MODE COLORS

\begin{abstract}

Recent mobility scaling research, using new data sources,
often relies on aggregated data alone. Hence, these studies face
difficulties characterizing the influence of factors
such as transportation mode on mobility patterns.

This paper attempts to complement this research by looking
at a category-rich mobility data set.  
In order to shed light on the impact of categories, as a case study,
 we use conventionally collected German mobility data.
In contrast to `check-in'-based data, our results are not biased by Euclidean distance approximations.

In our analysis, we show that aggregation can
hide crucial differences between trip length distributions, when
subdivided by categories. For example, we see that on an
urban scale (0 to $\sim15$ km), walking, versus driving, exhibits
a highly different scaling exponent, thus universality class.
Moreover, mode share and trip length are responsive to day-of-week and time-of-day.
For example, in Germany, although driving is relatively less frequent on Sundays than on Wednesdays, trips seem to be longer.

In addition, our work may shed new light on the debate
between distance-based and intervening-opportunity mechanisms
affecting mobility patterns, since mode may be chosen both
according to trip length and urban form.

    \vspace*{-.1cm}

\end{abstract}

\section{Introduction}

%"Big Data' is a novel and exciting set of data sources offering many opportunities that we are just beginning to tap.  However, one should not forget that more conventional sources still offer understanding of the same scaling phenomena being explored by these new types of information.

%APPLICATIONS -- NETWORKING, URBAN PLANNING, ETC.

It is important to understand mobility for a variety of reasons, including uncertainty reduction for allocation of resources such as communications and computing \cite{sousaaura} infrastructure usage, robustness and interdependence \cite{Vespignani2010}, wireless networking applications \cite{Griepp2013}, social network analysis \cite{Cho:2011:FMU:2020408.2020579}, intelligent transportation \cite{Silva2006}, economic development \cite{Helbing2004}, crisis response \cite{Chen2006, Helbing2000}, and large-scale energy consumption and CO$_2$ emissions \cite{Momoh2009, Townsend_2000, banister2011cities}, to name a few.

The data sources in recent papers in the `big' mobility-scaling literature have been dollar bill movements, mobile call-data records (CDRs), and geo-tagged social media such as Foursquare, Twitter, Gowalla, Facebook, and others \cite{Brockmann:2006uq, Gonzalez:2008uq, Noulas2011, Cho:2011:FMU:2020408.2020579}.  The defining characteristic of this big mobility data is not only its size, which can be smaller than some conventional sources -- the number of trips or movements in these `big' sources can range from $10^5$ to $10^9$ or broader -- but most often they are characterized by new forms of large-scale automatic data collection using `check-ins' (phone calls, tweets, etc.), for some purpose other than their eventual research use, at relatively low cost.

%%\begin{wrapfigure}{L}{0.38\textwidth}
%\begin{figure}[h!t]
%%%trim option's parameter order: left bottom right top
%\includegraphics[trim =9mm 55mm 9mm 51mm, clip, width=0.38\textwidth]{./city_mode_trip_length8.pdf}
%%\includegraphics[width=0.45\textwidth]{./city_mode_trip_length.pdf}
%\caption{Log-log trip length distributions (CCDF) of all trips starting in German cities with population $> 100,000$ by major transportation mode.}
%\label{fig:ccdf_city_trips_mode}
%\end{figure}
%%\end{wrapfigure}

The Where's George, CDRs, and social media contain either little or no categorical data about the trip or individual, or are limited due to privacy concerns.  Spatial resolution can vary from cell-tower radius ($\sim3$ km) to less than a few meters in the case of GPS-based social media  \cite{Brockmann:2006uq, Gonzalez:2008uq, Noulas2011, Cho:2011:FMU:2020408.2020579}. There are several challenges posed by these data sources, stemming from the large geographic- and time-scales, as well as the incidental sampling method, to be discussed below.

    \vspace*{-.1cm}

\subsection{Our Contributions}

Here, we are interested in the ability of conventionally-collected (`little') mobility data to contribute to scientific research on mobility patterns.  The availability of categorical information allows us to ask and address questions that are challenging using exclusively check-in mobility data.  Transportation mode, city size, and trip purpose are particularly helpful to shed light on mobility patterns at an urban scale.
%in understanding mobility patterns at an urban scale.  Details of our data will be discussed later, but
We say this data is conventional because it was collected as
%it was not collected in a low-cost automatic way for another purpose, but rather as
a large effort including survey design by experts as part of a series running over many years, with an intentional focus on understanding of mobility patterns.

%\begin{figure}[!ht]
%\subfloat[] {
%\includegraphics[scale=.3]{./_hwzweck_trip_length.pdf}
%\label{fig:zweck}}
%\vfill
%\subfloat[]{
%\includegraphics[scale=.3]{./_ktyp_trip_length.pdf}
%\label{fig:ktyp}}
%\caption{(\ref{fig:zweck})Trip length (CCDF) according to trip purpose. (\ref{fig:ktyp}) Trip length (CCDF) according to urban size.}
%\label{fig:purpose_area}
%\end{figure}

%\begin{figure}[h!t]
%\includegraphics[width = 0.4\textwidth]{./_hvm_diff_trip_length.pdf}
%\label{fig:diff_modes_trip_length}
%\caption{Detailed modes trip length CCDF  }
%\end{figure}

%This paper expands upon recent mobility pattern studies by others, but uses conventional, category-rich data.
Based on this
%conventional data, we shed new light on urban mobility and the impact of categorical data for scaling research. In particular,
we ask -- how can mode, trip purpose, and other categories further our understanding of mobility generally, and especially of urban mobility?
Also, how can we begin to address the challenges faced by big mobility research above?  For example, in contrast to most check-in displacements, which are inherently based on misleading Euclidean distances and may not correspond to actual trip start and end, we have reported lengths of the trips themselves.

Our main findings are:

\begin{itemize}

\item We argue that assuming that trips are i.i.d. is imprudent, and that categories matter in refining our understanding of mobility patterns.
Mode matters, helping to characterize mobility universality classes, both at the urban and inter-urban scales.  E.g. there are significant differences between walking, bicycling, and automobile driving trip length distributions.  In addition, looking at trip lengths rescaled by maximum length for each mode, there are significant distinguishable universal properties. % That is, even significantly different trips have similar dimensionless trip length exponents within certain ranges.

\item Even for trip lengths at and below the urban scale ($\sim 10$ km), mode differences are evident, a fact that is at odds with previous claims \cite{Noulas2011}.  On a related note, it seems that city population is not a strong determinant of mean trip length, with only a slight difference found in large vs. small cities.

\item Scaling of mobility confirms previous findings, when histograms of daily trips and overnight travel are taken together, yielding a scaling exponent of $1.44$ for trip lengths within Germany.

\item We show that other categories and dependencies are also important.  Trip lengths respond differently to different purposes, shopping and business, for example.  It is also evident that mobility is time-dependent.  E.g. trip lengths on Sunday vary from those on Wednesday.

 % We also address sampling time and the possible effect on apparent trip length distributions.  For example, bias may result if time is not taken into account.

%\item Finally, there seems to be some universality in scaling of trip lengths with respect to maximum trip length by different modes.  That is, even significantly different trips have similar dimensionless trip lengths.

%\item Finally, we look at trip purpose for particular modes (walking and bicycling) and show that they respond differently to purpose.

\end{itemize}

More broadly, we see that:
Scaling in response to categories hints at the existence of different universality classes in mobility patterns;
purpose, along with mode, may give us insight into how to form a bridge between distance and intervening opportunity arguments for trip length form in cities;
since this dataset is less prone to sampling error, we believe it can also offer the ability to understand trip length changes in time, and helps elucidate some of the factors that are averaged together when using large-scale check-in data.

% start page numbering on page 2
\thispagestyle{plain}
\pagestyle{plain}

\section{Related Work and Challenges}

\begin{figure}[!ht]
    \vspace*{-.2cm}
    \subfloat[]{%
    \includegraphics[trim =9mm 55mm 9mm 51mm, clip, width=0.24\textwidth]{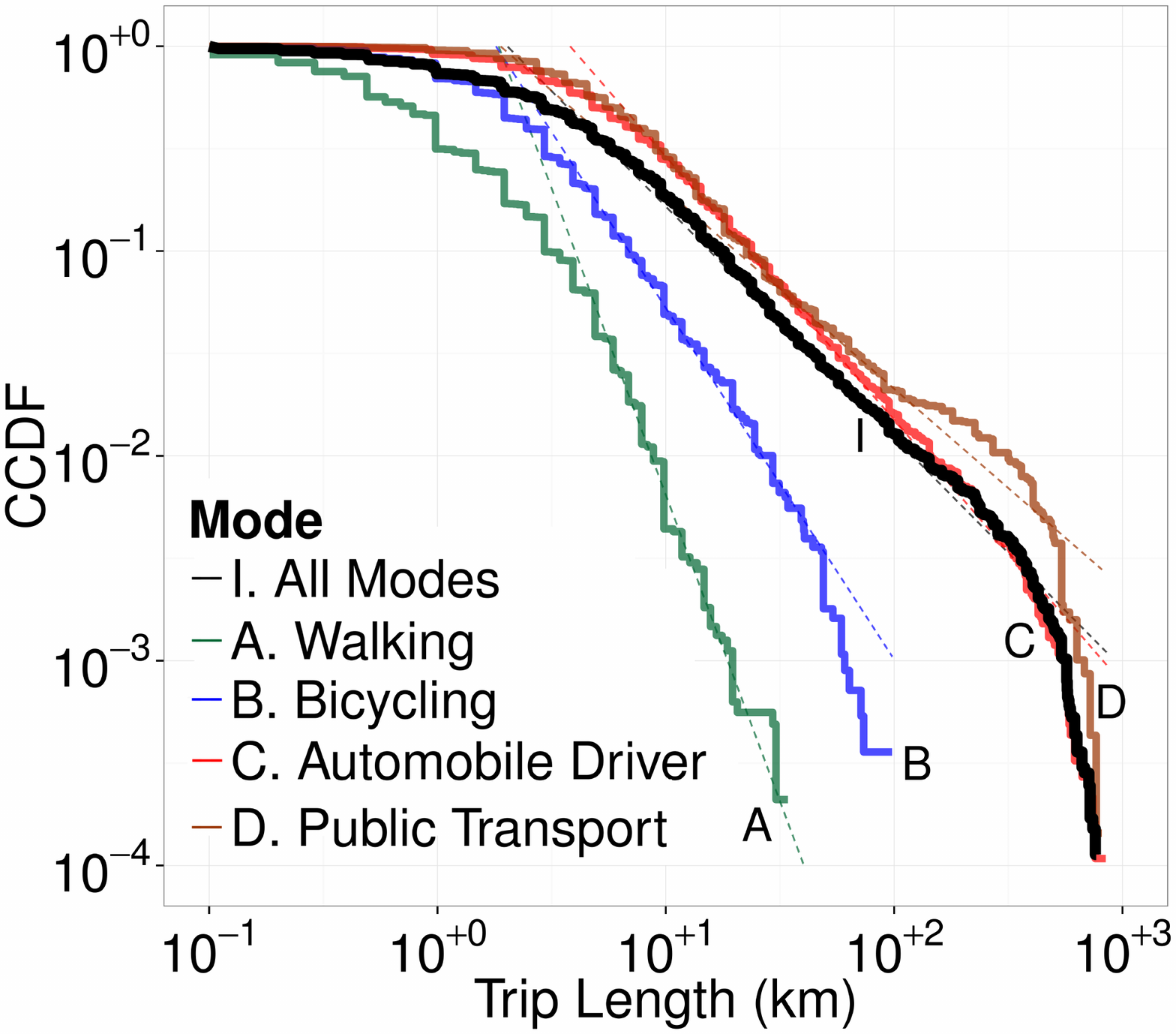}
\label{fig:ccdf_city_trips_mode}
    }
    \subfloat[]{%
 %%trim option's parameter order: left bottom right top
\includegraphics[trim =9mm 55mm 9mm 51mm, clip, width=0.24\textwidth]{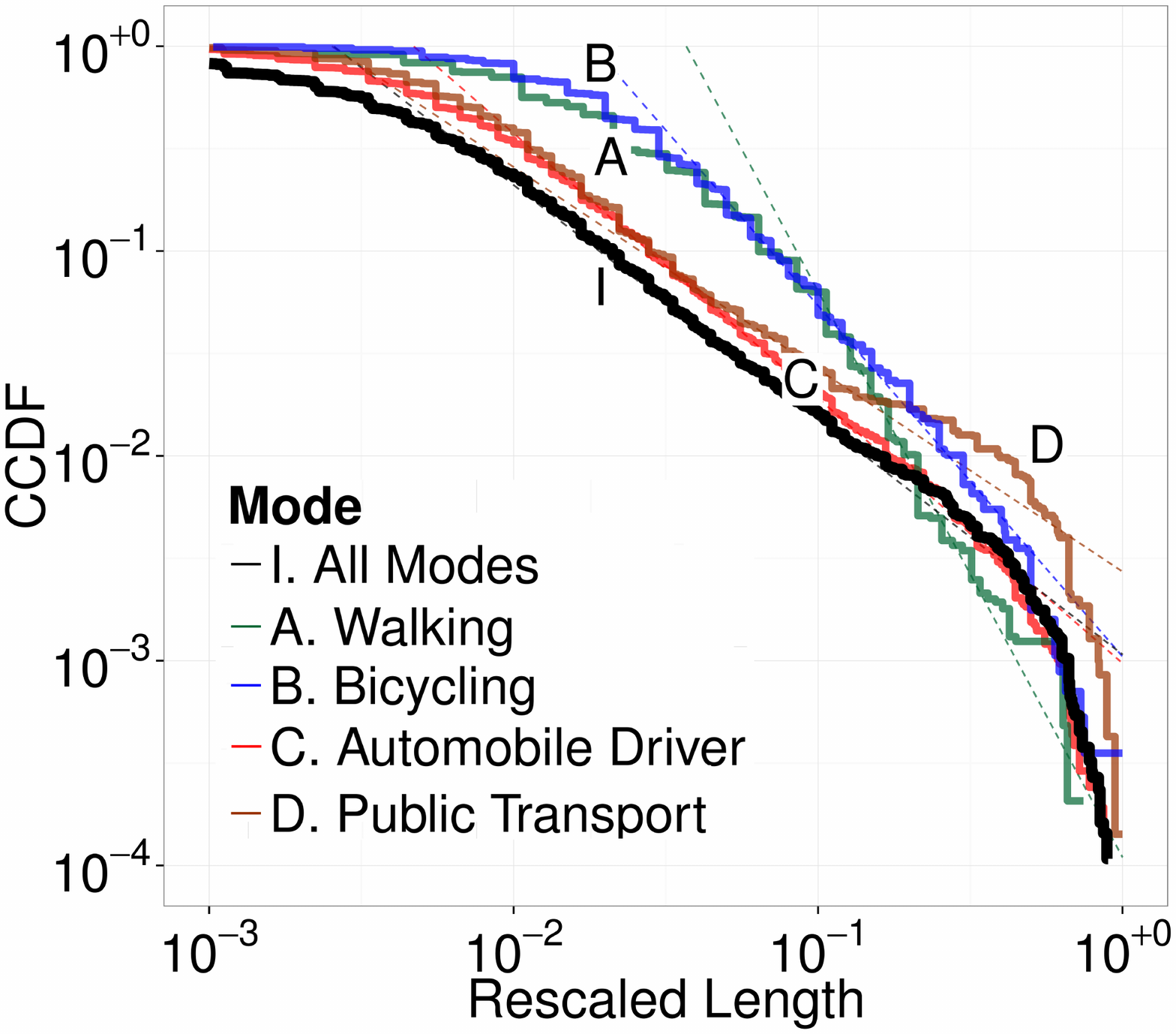}
\label{fig:rescaled_trips}
    }
    \caption{ (\ref{fig:ccdf_city_trips_mode}) Trip length distributions (CCDF) of all trips starting in German cities with population $> 100,000$ by major transportation mode.  (\ref{fig:rescaled_trips}) CCDF of trip lengths rescaled by maximum trip length $(\ell/\ell_{max})$ for each respective mode.}
    \label{fig:mode}	
    \end{figure}

There has recently been progress characterizing scaling of long-distance trips ($\sim10^2$ to $\sim10^4$ km), fitting them with power law having scaling exponent ranging between $1.50$ and $1.75$ \cite{Noulas2011, Gonzalez:2008uq, Brockmann:2006uq}.  Since at the longer scale fewer (motorized) modes dominate mobility, and they are somewhat clustered compared to non-motorized modes (See Fig.~\ref{fig:ccdf_city_trips_mode} and Table. \ref{table:fits}), it is not surprising that these trips are easier to characterize.  There has been some success modelling these long trips and attempting to determine major mechanisms driving them.   These mechanisms -- some also found in past research using conventional data -- are based on distance (Random Walks and Levy Flights) \cite{Brockmann:2006uq, Gonzalez:2008uq}, `intervening opportunity' (place density) \cite{simini_2012, Stouffer1940}, along with social networks \cite{Cho:2011:FMU:2020408.2020579}, and others \cite{song2010modelling}.
    
Longer trips are very different from spontaneous, inexpensive, dense infrastructure, dense location, urban-scale mobility, occurring at distances less than or close to $10^1$ km \cite{Noulas2011, Scheiner2010}.  Mobility research has been facing an `urban challenge', due to: A. Difficulties fitting shorter distance trips, or the necessity of using distance transformations \cite{Noulas2011, Gonzalez:2008uq}, B. the limits of spatial resolution (e.g. in CDRs) \cite{Gonzalez:2008uq}, and C. because heavy-tail research is focused exactly on the {\it tail} of distributions, due to the systemic and mathematical property of power law scaling -- to break down at small data values \cite{Newman2009}.

% the averaging of trips over all categories, purposes, urban densities, times, and other factors;

% For urban scale trips, research has used data transformations \citep{Gonzalez:2008uq} or explored other mechanisms \cite{Brockmann:2006uq,Song2010,Cho:2011:FMU:2020408.2020579,Noulas2011} (place density, social networks, exploration of new destinations, rate of return) or just focused on analysis of long-range patterns  .

%It is difficult to characterize the entirety of big mobility data research.  Most generally, these works have explored cities-as-particle-systems, attempting to determine major mechanisms driving mobility patterns.   These mechanisms -- some also found in past research using conventional data -- are trip distance \cite{Brockmann:2006uq} \cite{Gonzalez:2008uq}, place density and `intervening opportunity' mechanisms \cite{simini_2012, Noulas2011, Stouffer1940}, along with social networks \cite{Cho:2011:FMU:2020408.2020579}, and others \cite{Song2010}.

Recently, this work has attempted to address the apparent debate between two schools of thought about mechanisms influencing mobility patterns on the urban scale:  A. distance-based mechanisms \cite{Gonzalez:2008uq, Song2010}, versus B. intervening opportunity \cite{Stouffer1940, Noulas2011}.  Basically, their question is: Is there some inherent property of human behavior -- purely related to distance -- that leads to heavy-tailed trip-length distributions, or are these trip length patterns driven more by urban form, as seen in the density of `places'?

This mobility research faces some significant challenges:

First, these check-in sources do not usually contain very much ancillary categorical information about a trip such as mode, weather, purpose, number of passengers, etc.  Thus, if one wants to know the effect of external factors, one may be limited by resources to determine all of them accurately~\cite{Reddy2010}.

Second is sampling bias.  Between check-ins, it may be impossible to know actual travel patterns, and check-in rates may not be independent of factors (such as mode) that affect these patterns.  Trips may not begin and end at check-ins, and  very rarely follow a linear path -- the terms `travel' and `displacement' are intermingled \cite{Brockmann:2006uq, Gonzalez:2008uq, Noulas2011}, which may be appropriate at distances mostly traversed by air, but certainly can be misleading at urban scales.  Shorter trips length measurements may be more sensitive to these inaccuracies.

With one or two significant exceptions, existing mobility scaling research seems to implicitly assume that `mean field', random, independent characteristics apply -- due to the large data size -- and that these approximations are sufficient to account for sampling bias \cite{Brockmann:2006uq, Gonzalez:2008uq, Noulas2011}, so that check-in displacements are assumed to reflect actual displacements or trip lengths.  %It is reasonable to question this assumption.  (For example, imagine CDR or social media check-in rates while walking, versus driving.  It is highly unlikely that they are identical, and therefore sampling may bias apparent trip distance distributions.)

%stationarity or
Finally, related to sampling bias is stability of mobility patterns over time.  There is no question that mode share and trip length change, and that this needs to be considered.% \cite{Gonzalez:2008uq}.  %Due to the often sparse, large-scale nature of Big Data and the need for data density to make `mean field' arguments, temporal stability seems to pose a challenge, since subsetting within time ranges may result in data points that are too sparse to make confident conclusions.  Conventional data sources will of course also contain sampling bias, but we might ask whether this bias significantly affects trip length statistics, and whether time dependencies can be accounted for.

    \vspace*{-.1cm}

\section{Data and Methodology}

This work is based on the Mobility in Germany 2008 (MID 2008) survey data set, which was collected and is maintained by the Infas Institute for Applied Social Science Research and the German Aerospace Center (DLR), with the main survey between the dates of February 2008 and March 2009.  The final survey involved 25,922 Households, 60,713 Individuals, 193,290 Trips and 36,182 Travel events.  `Trips' describe daily journeys, where a return journey was counted as a separate trip, while `travel' data describe mobility that included an overnight stay \cite{Follmer2008a}.

MID 2008 was designed carefully, as a continuation of the West German Kontiv surveys in 1976, 1982 and 1989, and MID 2002.  It included a pre-survey, pretest, and used a mixed methodology combination of computer-aided telephone interview (CATI), online, and mail surveys in order to avoid bias and maintain continuity with past surveys.  Querying a large number of households from different federal states, it was the largest household survey apart from the official German microscensus.

The trip lengths $(\ell)$ in our data correspond to the {\it actual} traveled distances, reported by subjects.  Hence, in contrast to check-in data, we do not have to approximate trip lengths by Euclidean displacements $(\Delta r)$ \cite{Brockmann:2006uq, Gonzalez:2008uq, Noulas2011}, which may introduce a bias to the scaling exponent, especially for short trips.  This is particularly interesting, since our data features a high resolution, recorded down to the 100m scale.

%Here we use {\it trip lengths} ($\ell$) instead of {\it displacements} ($\Delta r$), since we have reports of distance traveled.  These distances are approximate, recorded down to 100 meter resolution.

If not otherwise stated, lengths shown are for trips only, not travel, and trips are counted over the entire measurement period.  Categorical information describe trip origination and mode describes the main transportation mode for a trip.  We define urban trips as those starting in a city (pop. $> 100,000$), and other categorical information is stated explicitly.  `All modes' is composed of a weighted average of walking, bicycling, automobile drivers, automobile passengers and public transportation trips.  We have removed the automobile passenger mode from figures for ease of visibility, but note that its scaling and statistical characteristics are similar to those of public transportation (Table \ref{table:fits}).
    \vspace*{-.1cm}

    \vspace*{-.1cm}

We simply use {\it best-fit} power law scaling exponents ($\alpha$) to give a sense of relative scaling in what are visibly truncated heavy-tailed distributions, not as claim to fit.  Power laws are of the form $p(\ell) = C \ell^{-\alpha},$
for normalization constant $C$, trip length $\ell$, scaling exponent $\alpha$, and $\ell > \ell_0,$ the minimum fit trip length.  Here we have shown trip lengths as log-log CCDFs, $p(L > \ell)$, as is common in scaling literature \cite{Newman2009}.

Statistical fitting was carried out by a method that uses maximum likelihood estimators and Kolmogorov-Smirnov statistics to fit data with a power law. (See \cite{Newman2009}.) %For goodness of fit, it then tests statistics of the best-fit CDF against best-fit CDFs of synthetic data sets \cite{Newman2009}.

\section{The Importance of Categories}

\subsection{Mode Matters for Mobility Scaling}
\label{sec:mode}

\begin{table}[!ht]%\footnotesize
    \vspace*{-.2cm}

\centering
\begin{tabular}{|l|c|c|c|c|c|c|}
\hline
Mode & Count & $\alpha$ & $\ell_0$ (km) & $\bar{\ell}$ (km) & $\sigma^2$\\
\hline
I. All Modes & 52973 & 2.13 & 29.40 & 9.99 & 1313.79\\
\hline
A. Walk & 14303 & 3.99 & 6.37 & 1.37 & 3.77\\
\hline
B. Bicycle & 5581 & 2.72 & 6.37 & 3.47 & 30.06\\
\hline
C. Auto. Driver & 18484 & 2.29 & 39.90 & 13.06 & 1331.84\\
\hline
D. Public Trans. & 6944 & 1.97 & 27.98 & 16.34 & 2875.92\\
\hline
Auto. Passenger & 7658 & 2.00 & 24.32 & 17.69 & 2949.11\\
\hline
\end{tabular}
\caption{Sample size, best-fit scaling exponent $\alpha$, beginning of fit $\ell_0$, and moments -- mean trip length $\bar{\ell}$ and variance $\sigma^2$ for the major modes.}
\label{table:fits}
\end{table}
\vspace*{-.1cm}

Check-in-based mobility data research must average together displacements of substantially different modes.  With our conventional categorical data we can distinguish between modes, seeing a visible discrepancy in scaling between walking, bicycling, automobile drivers, and those using public transport (Fig.~\ref{fig:ccdf_city_trips_mode}).  For comparison of the scaling, best-fit power laws are drawn, with corresponding exponents shown in Table \ref{table:fits}.
%We see heavy-tailed trip lengths for each mode, and observe that one should not expect to fit a trip length distribution resulting from the aggregation of modes as varied as walking and driving.
%As expected, walking and bicycling exhibit much shorter trip length than other modes.
With $\alpha > 3$ for walking (A), the first two moments -- mean and variance -- are defined.  For bicycling and driving (B, C), with $2 < \alpha < 3$, the mean is defined but variance diverges.  For public transport (D), with $1 < \alpha <2$, neither the mean nor variance is defined \cite{newman2005power}.

\begin{figure*}[h!t]
\subfloat[]{
\includegraphics[trim =9mm 55mm 9mm 45mm, clip, width=0.32\textwidth]{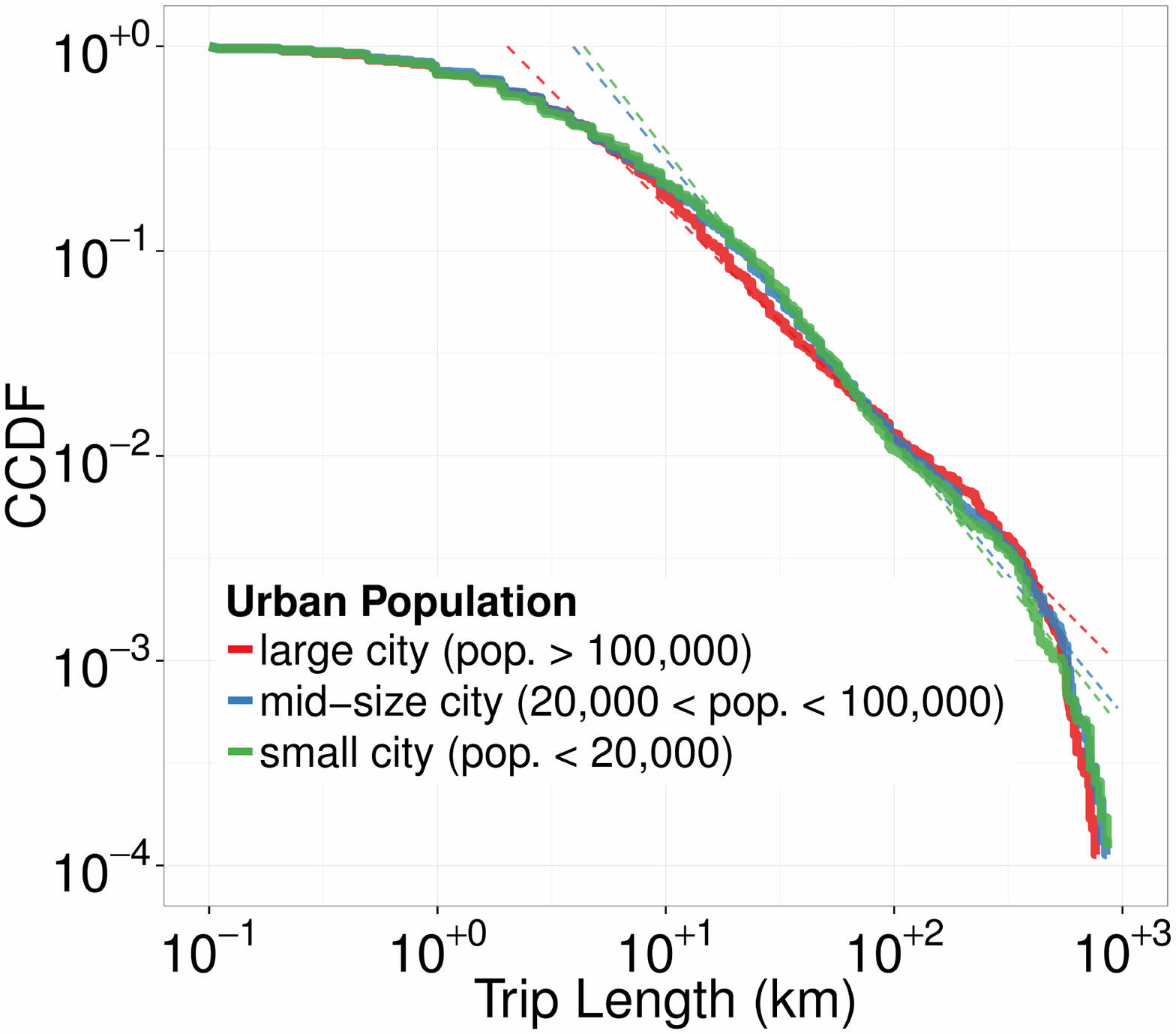}
\label{fig:municipality_type}}
\subfloat[]{
\includegraphics[trim =9mm 55mm 9mm 45mm, clip, width=0.32\textwidth]{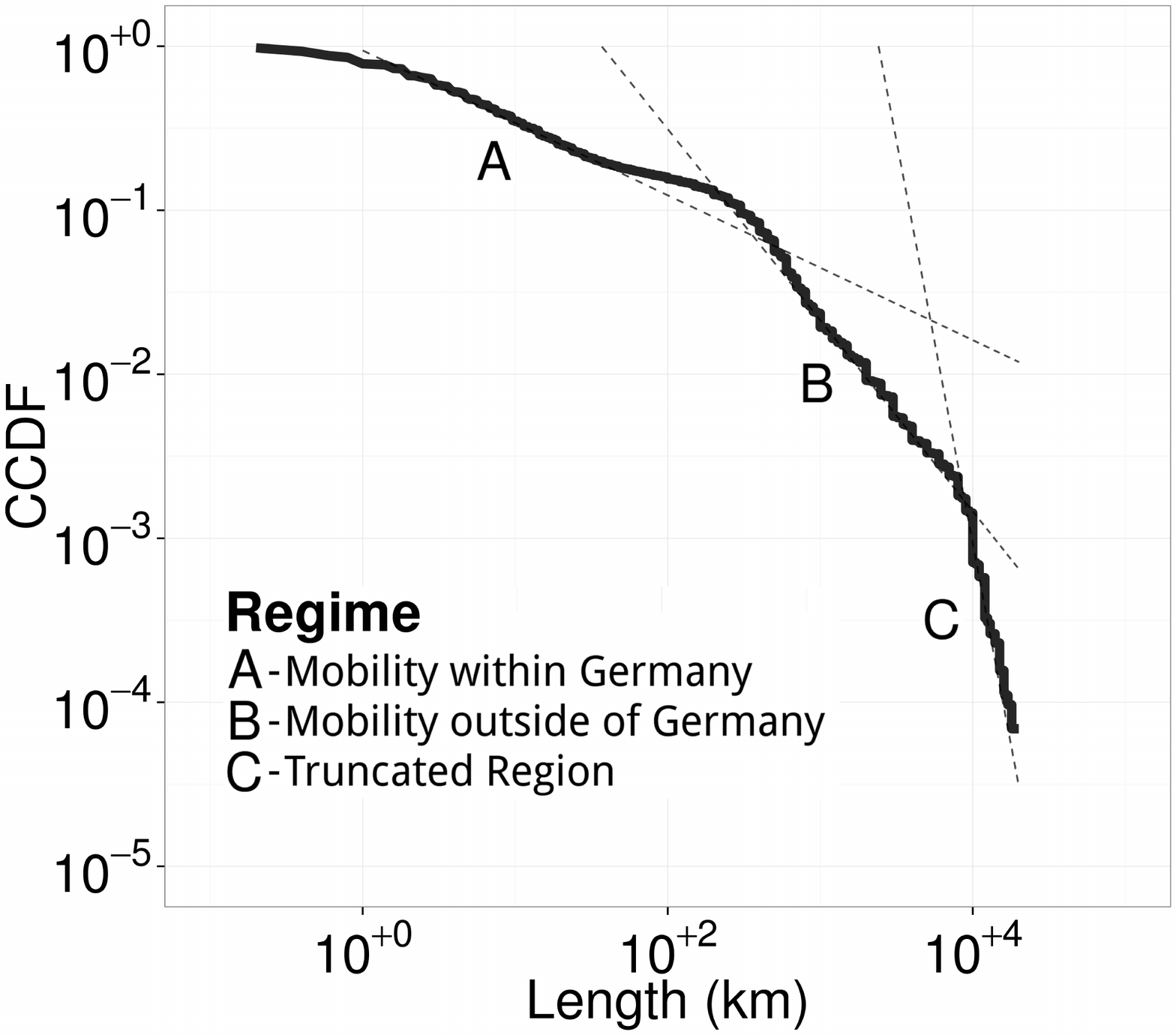}
\label{fig:travels_and_trips}}
\subfloat[]{
\includegraphics[trim =9mm 55mm 9mm 45mm, clip, width=0.32\textwidth]{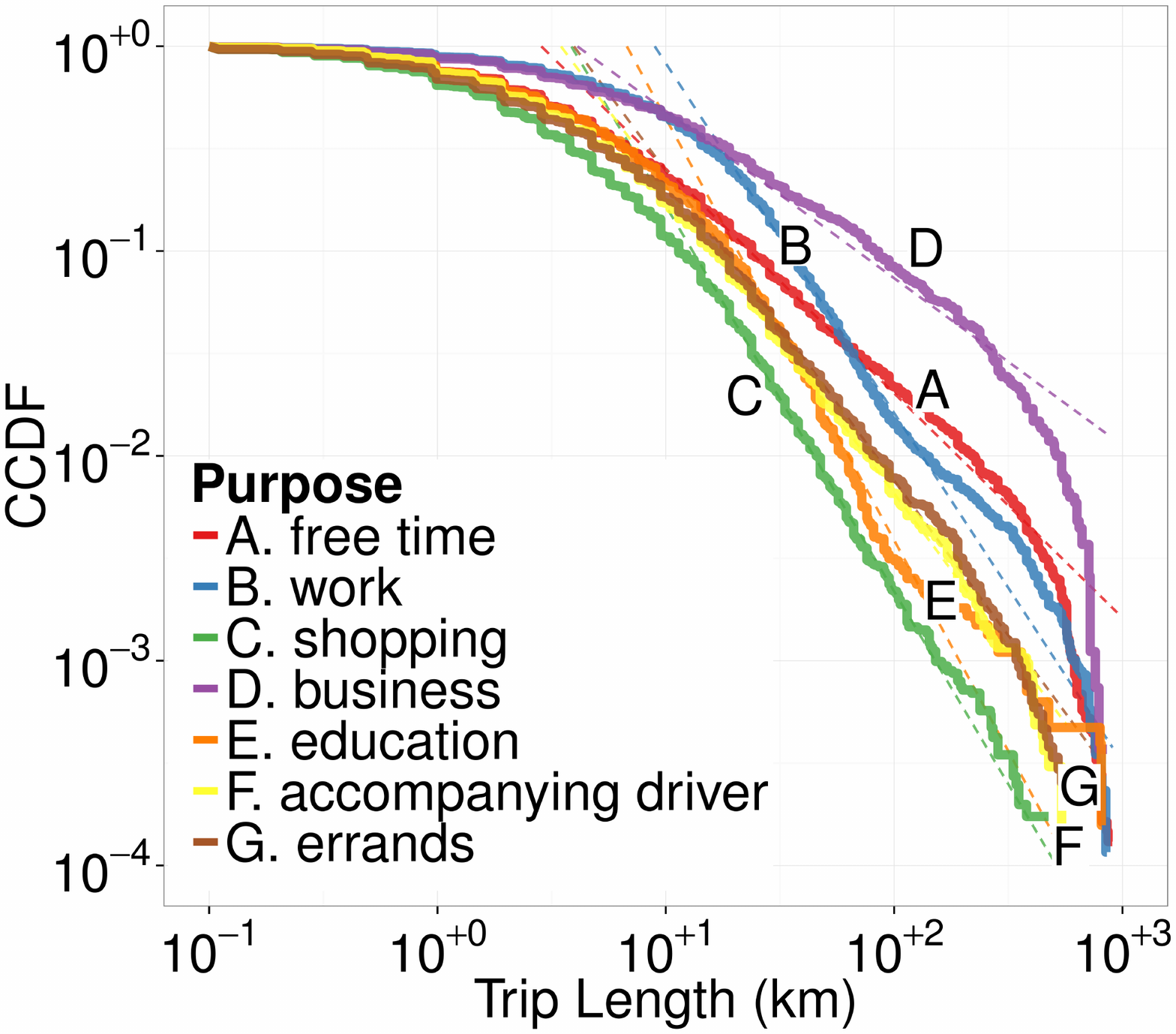}
\label{fig:purpose}}
\caption{CCDFs of (\ref{fig:municipality_type}) trip length according to urban population, (\ref{fig:travels_and_trips}) daily trip and overnight travel lengths originating in Germany, taken together,  (\ref{fig:purpose}) trip length according to purpose.}
\end{figure*}
%\vspace*{-.1cm}

%\begin{table}[!ht]%\footnotesize
%\centering
%\begin{tabular}{|l|c|c|c|c|c|c|}
%\hline
%Mode & $\alpha$ & $\ell_0$ & $\bar{\ell}$ & $\sigma^2$\\
%\hline
%I. All Modes  & 2.15 & 0.03 & 0.01 & 1.79$\times 10^{-3}$\\
%\hline
%A. Walking &  3.76 & 0.14 & 0.03 & 2.03$\times 10^{-3}$\\
%\hline
%B. Bicycling & 2.71 & 0.06 & 0.04 & 3.11$\times 10^{-3}$\\
%\hline
%C. Auto. Driver & 2.29 & 0.04 & 0.02 & 1.76$\times 10^{-3}$\\
%\hline
%D. Public Trans. & 1.99 & 0.03 & 0.02 & 4.55$\times 10^{-3}$\\
%\hline
%Auto. Passenger & 2.00 & 0.03 & 0.02 & 4.13$\times 10^{-3}$\\
%\hline
%\end{tabular}
%\caption{Best-fit scaling exponent, beginning of fit, mean trip length (rescaled by maximum) and variance.}
%\label{table:rescaled_fits}
%\end{table}

Scaling exponent ($\alpha$) and mean trip length ($\bar{\ell}$) for walking and bicycling (Table \ref{table:fits}: A,B), both non-motorized, differ greatly from that of motorized modes- automobile driving and public transportation usage (C, D), as one might expect.  We also note that walking and bicycling have exponents differing by more than one, and that the exponent for walking, nearly 4, implies that it behaves quite differently than other modes.

Note that trip lengths, representing daily trips originating within Germany, are truncated at approximately the diameter of Germany\footnote{Using a simplifying approximation of a disc, we calculate $\log_{10}$(diameter) $= \log_{10} (2\sqrt{\frac{A}{\pi}}) \approx 2.83$, where $A = 357,021$ km$^2$, Germany's square area.} (674 km $\approx 10^{2.83}$ km, see Fig.~\ref{fig:ccdf_city_trips_mode}).  This is confirmed by intuition, since it is perhaps less likely that trips beginning in Germany, not including an overnight stay, will end in another country.

%\subsection{Rescaled Distance}

Rescaling trip lengths by the maximum trip length for each respective mode, we also observe that certain modes have somewhat similar heavy tails (Fig.~\ref{fig:rescaled_trips}), again suggesting distinct universality classes, and thus some mechanism at work causing these differences.   Between $\sim 10^{-3}$ and $\sim 10^{-1}$ of maximum trip length, trips seem clustered into two groups by scaling, non-motorized -- walking and bicycling, and motorized -- auto. driving and public transport.  From $10^{-1}$ to $10^{0}$ of max. trip length, scaling for the various modes seems to diverge.

Generally, correlation of mode with trip length scaling has considerable implications for human systems such as cities.  A small change in a mode's scaling exponent can imply a large difference in total trips of a certain length, and therefore total energy.  Mode share also implies a significantly different energy consumption budget.  (E.g. walking vs. automobile modes.)  Since these statistics describe system characteristics of large-scale random processes -- sometimes called `urban metabolism' \cite{Townsend_2000, banister2011cities, wolman1965metabolism, west2004life} -- and therefore substantial amounts of energy and $CO_2$ emissions, they are very important to understand.

%We also note that mode choice implies a different energy consumption budget.  Trivially, walking, and automobile modes have significantly different energy characteristics.  This has large economic and climate change implications, when we consider the universality of these patterns over countries or continents.

\subsection{Urban Mobility Patterns}
    \vspace*{-.2cm}

\begin{table}[!ht]\footnotesize
\centering
\begin{tabular}{|c||c|c|c|c|c|}
\hline
Length & Mode \% & • & • & • & • \\
\hline
%• & Walking & Bicycling & Driving & Pub. Trans. & Auto. Pass.\\
%\hline
%intra-urban & 28.30 & 11.90 & 38.06 & 14.73 \\
%\hline
%inter-urban & 00.31 & 1.83 & 58.60 & 14.57 & 24.68 \\
• & walking & bicycling & auto. & public & auto. \\
• & • & • & driving & trans. & passenger  \\
\hline
intra-urban  & 28.30 & 11.91  & 38.06 & 6.99 & 14.73 \\
\hline
inter-urban & 0.31 & 1.83  & 58.60 & 14.57 & 24.68  \\
\hline
%mode" & "Fahrrad" & "keine Angabe" & "MIV (Fahrer)" & "MIV (Mitfahrer)" & "PV" & "zu Fu"\\
%\hline
%"intra.mode.shares" & 11.9092126284742 & 0.00388090787371526 & 38.0632976074203 & 14.73386674256 & 6.98628099066642 & 28.3034611230054\\
%\hline
%"inter.mode.shares" & 1.83161840104664 & 0.00707188571832679 & 58.6047169477741 & 24.6773452141013 & 14.5680845797532 & 0.311162971606379\\
%\hline
\end{tabular}
\caption{Mode share $(\%)$ for intra-urban ($< 10^{1.17}$ km) and inter-urban trips ($\ge 10^{1.17}$ km) for large cities (pop. $> 100,000$) in Germany.}
\label{table:short_vs_long_mode_share}
\end{table}
\vspace*{-.2cm}

For Germany's 76 cities with over 100,000 population, the average area is 174.02 km$^2$ \cite{Office}.   Using a similar approximation as for Germany$^1$, this yields an urban diameter of 14.89 km  ($\approx 10^{1.17}$ km).

Mode is therefore also revealing about urban scale mobility, since we can now use trip length statistics separated by mode to distinguish between patterns near and below this scale (Fig.~\ref{fig:ccdf_city_trips_mode}).
For intra-urban trip lengths below the urban diameter, non-motorized modes contribute significantly to trip statistics (Table~\ref{table:short_vs_long_mode_share})  At the inter-urban scale, trip statistics are mostly the result of motorized modes, as expected.
%\footnote{These intra-urban mode shares are largely confirmed by third-party statistics collected for Berlin \cite{berlin_mobility}. }.  

It is important to note these scaling differences, especially in intra-urban region.  Here, averaging together all of these modes (`all modes') is essentially averaging the heads of some trip length distributions together with the tails of others (See $\ell_0$ and $\bar{\ell}$, Table \ref{table:fits} and Fig. \ref{fig:ccdf_city_trips_mode}), and thereby aggregating the results of processes belonging to significantly distinct universality classes.  It is therefore not surprising that urban scale mobility patterns have posed a challenge to those using aggregated check-in data.

As noted above, the non-motorized versus motorized modes each seem to be the product of some unique mobility process at the urban scale -- since both their absolute and rescaled trip length distributions stand apart (Figs. \ref{fig:ccdf_city_trips_mode} and \ref{fig:rescaled_trips}).  These largely different exponents imply that trips by certain modes are caused by different processes and system characteristics, belonging to distinct universality classes -- plausible when comparing these groups of modes.  This also suggests that we may be able to consider modes as making up separate phases of the underlying process of mobility \cite{newman2005power, Mitzenmacher2004, Stanley1999}.  Also, due to the different form of rescaled trip lengths for non-motorized modes -- perhaps exponential -- this is interesting to consider in the context of mobility behavior of other organisms \cite{ramos2004levy}.  Thus with this information, we can begin to investigate causal mechanisms more carefully.

It seems that mode allows us to describe trip lengths primarily by their scaling exponent within the intra-urban region, perhaps down as far as $\ell_0$ = 6.37 km ($10^{.8}$ km) (Table \ref{table:fits}).  However, below that distance other factors may be at work, and the behavior may be better described primarily by something other than scaling with respect to the mode category.

\begin{table}[!ht]\footnotesize
\centering
\begin{tabular}{|c|c|c|c|c|c|}
\hline
Urban Population & Count &$\alpha$ & $\ell_0$ (km) & $\bar{\ell}$ (km) & $\sigma^2$ \\
\hline
small ($< 20k$) & 23433 & 2.41 & 43.32 & 10.52 & 1202.28 \\
\hline
medium ($20k$-$100k$) & 53038 & 2.35 &  30.38 & 10.62 &  1329.72 \\
\hline
large ($> 100k$) & 53011 & 2.13 & 29.40 & 9.99 &  1312.92 \\
\hline
\end{tabular}
\caption{Sample size (Count), best-fit scaling exponent ($\alpha$), beginning of fit ($\ell_0$), mean trip length ($\bar{\ell}$) and variance ($\sigma^2$) according to city population.}
\label{table:trip_length_vs_pop}
\end{table}
%\vspace*{-.1cm}

On a related note, trip lengths seem related to urban population, but not strongly (Fig.~\ref{fig:municipality_type} and Table~\ref{table:trip_length_vs_pop}), confirming other results \cite{Scheiner2010}.  For example, there is a small difference between mean trip lengths $(\bar{\ell})$ in low-population rural municipalities versus larger urban populations.  It therefore seems further investigation is needed to determine whether mean trip length scales allometrically with city population alone, as has been found for other urban parameters \cite{Helbing2007}.

Also, this indeterminate response by trip length to city population may support previous results about the independence of trip length and {\it city area} \cite{Noulas2011}, but since the Pearson correlation of urban population and area in Germany is not high $(r = 0.51)$ \cite{Office}, this cannot yet be confirmed.
	
\subsection{Trips taken together with overnight travel confirm previous findings}
 \begin{table}[!ht]\footnotesize
\centering
\begin{tabular}{|c|c|c|c|c|c|c|}
\hline
Regime & Count & $\alpha$ & $\ell_0$ (km) & $\bar{\ell}$ (km) & $\sigma^2$\\
\hline
%cat.val & num.trips & alpha & x.min & av & v1\\
%\hline
A & 209,045 & 1.44 & 1.81 & 48.97 & 14,727.00\\
\hline
B & 8,055 & 2.17 & 816.00 & 1,670.36 & 2,172,741.49\\
\hline
C & 380 & 5.91 & 11,000.00 & 11,312.92 & 7,047,781.70\\
\hline
\end{tabular}
\caption{Count, $\alpha$, $\ell_0$, $\bar{\ell}$ and $\sigma^2$ for the three distance regimes of trips and travel taken together.}
\label{table:trips_travel_fits}
\end{table}
\vspace*{-.1cm}

Furthermore, if we take daily trips and overnight travel together (Fig.~\ref{fig:travels_and_trips}), there seem to be three regimes:  (A) Within Germany, (B) outside of Germany, and (C) near the maximum distance that can be traveled from Germany to the other side of the world.	
		
For trips within Germany (Regime A), our best-fit gives us a scaling exponent of $\alpha = 1.44$, which is proximate to that found for Foursquare data ($\alpha = 1.50$) \cite{Noulas2011}, and for the Where's George data ($\alpha = 1.59$) \cite{Brockmann:2006uq}, though not as near to that found using call data records ($\alpha = 1.75$) \cite{Gonzalez:2008uq}.  Similar to trips without overnight travel (Fig.~\ref{fig:ccdf_city_trips_mode}), this is truncated by the diameter of Germany ($\sim 10^{2.83}$ km).

For longer trips outside of Germany (Regime B), our best-fit result is quite different from others ($\alpha = 2.24$).   However, big mobility data sources can include trips from all possible origins.   Since our data was collected differently and only includes journeys originating within Germany, is not surprising that we see a marked decrease in the number of trips of this length.  This second regime is truncated at roughly the distance of the furthest significant travel destination, Southeast Asia.  (E.g. The flying distance from Germany to Thailand is approximately 8667 km $\approx 10^{3.94} $ km.)  This truncation seems to agree with 2008 travel planning statistics, which show that few journeys ($ < 1\%$) were planned farther than Asia \cite{adac_reisemonitor_2008}.

\subsection{Distance-based and intervening opportunity arguments}

\begin{table}[!ht]%\footnotesize
\centering
\begin{tabular}{|l|c|c|c|c|c|c|}
\hline
Purpose & Count & $\alpha$ & $\ell_0$ (km) & $\bar{\ell}$ (km) & $\sigma^2$\\
\hline
%cat.val & num.trips & alpha & x.min & av & v1
education & 12704 & 3.06 & 31.07 & 8.15 & 574.29\\
\hline
shopping & 40322 & 2.88 & 35.15 & 5.19 & 196.73\\
\hline
work & 25808 & 2.71 & 38.95 & 17.40 & 1654.51\\
\hline
errands & 23716 & 2.51 & 45.13 & 8.06 & 593.81\\
\hline
accompanying driver & 16447 & 2.50 & 32.30 & 7.74 & 476.70\\
\hline
free time & 61152 & 2.10 & 30.38 & 13.55 & 2209.65\\
\hline
business & 2706 & 1.82 & 12.35 & 36.58 & 8011.00\\
\hline
\end{tabular}
\caption{Count, $\alpha$, $\ell_0$, $\bar{\ell}$ and $\sigma^2$ for trips by purpose.}
\label{table:purpose_fits}
\end{table}
\vspace*{-.1cm}

This mode information lets us address the central premise of a previous work, which suggested that trip length patterns cannot be distinguished at an urban scale \cite{Noulas2011}.  These authors then went on to give convincing arguments that `intervening opportunity' -- using rank-distance of place -- can largely explain urban trip length patterns, rather than purely distance-based mechanisms.  

Here, however, we have seen that trip lengths according to mode are distinguishable at this scale, lending credence to distance-based mechanisms.  Our evidence does not necessarily contradict their conclusions, but rather allows us to hypothesize that mode, together with trip purpose -- both obviously strongly correlated with trip length (Figs. \ref{fig:mode} and \ref{fig:purpose}) -- can help elucidate the debate between these apparently disparate schools of thought.  The distinct response of trip length to purpose (Fig. \ref{fig:purpose}) seems to support this line of thinking, since by necessity trip length according to purpose must respond to urban form (density and location of schools or grocery stores, for example).  Another work analyzing earlier versions of our data set has also suggested that trip distance is a function of facility location (urban form), which then determines mode \cite{Scheiner2010}.  Certainly, further work is needed, such as multivariate analysis and clustering.

\section{The Influence of Time}
 
 \begin{figure}[!ht]
%     \vspace*{-.2cm}

    \subfloat[]{%
    %    %%trim option's parameter order: left bottom right top
      \includegraphics[trim =9mm 55mm 7mm 45mm, clip, width=0.24\textwidth]{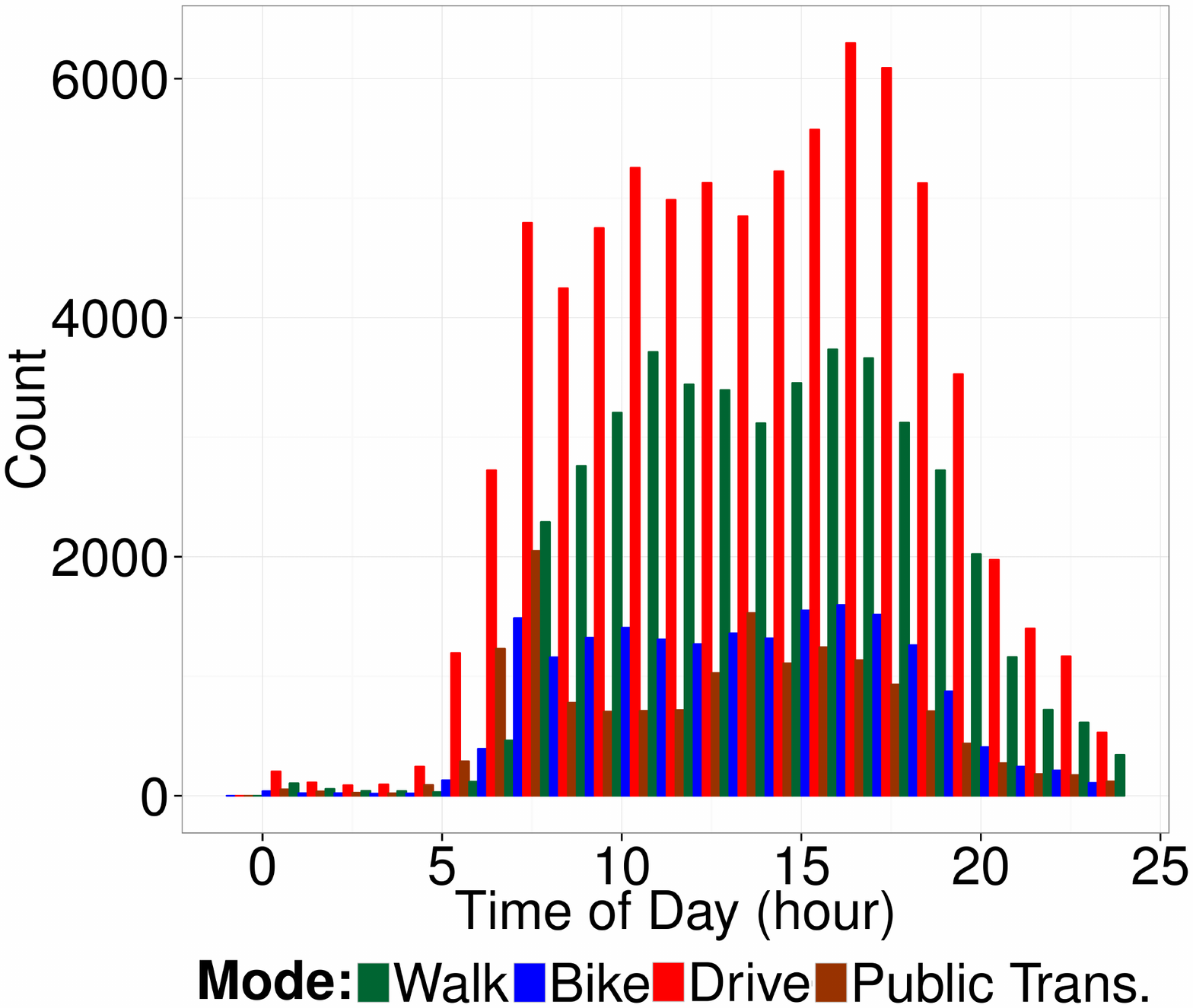}
      \label{fig:mode_share_hourly}
    }
    \subfloat[]{%
    \includegraphics[trim =9mm 55mm 7mm 45mm, clip, width=0.24\textwidth]{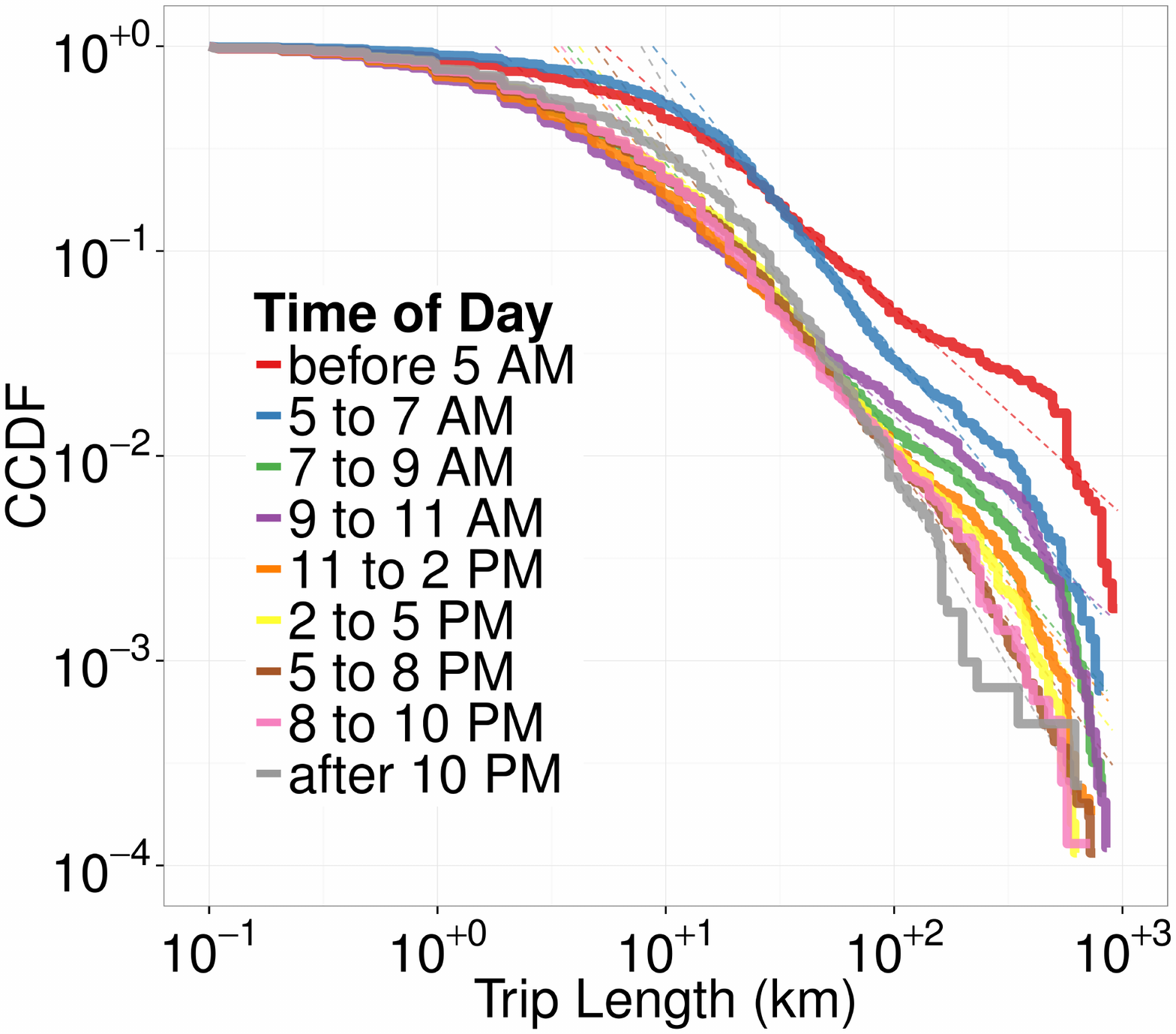}
      \label{fig:hourly_trip_lengths}
    }
    \vfill
    \subfloat[]{%
      \includegraphics[trim =9mm 55mm 9mm 45mm, clip, width=0.24\textwidth]{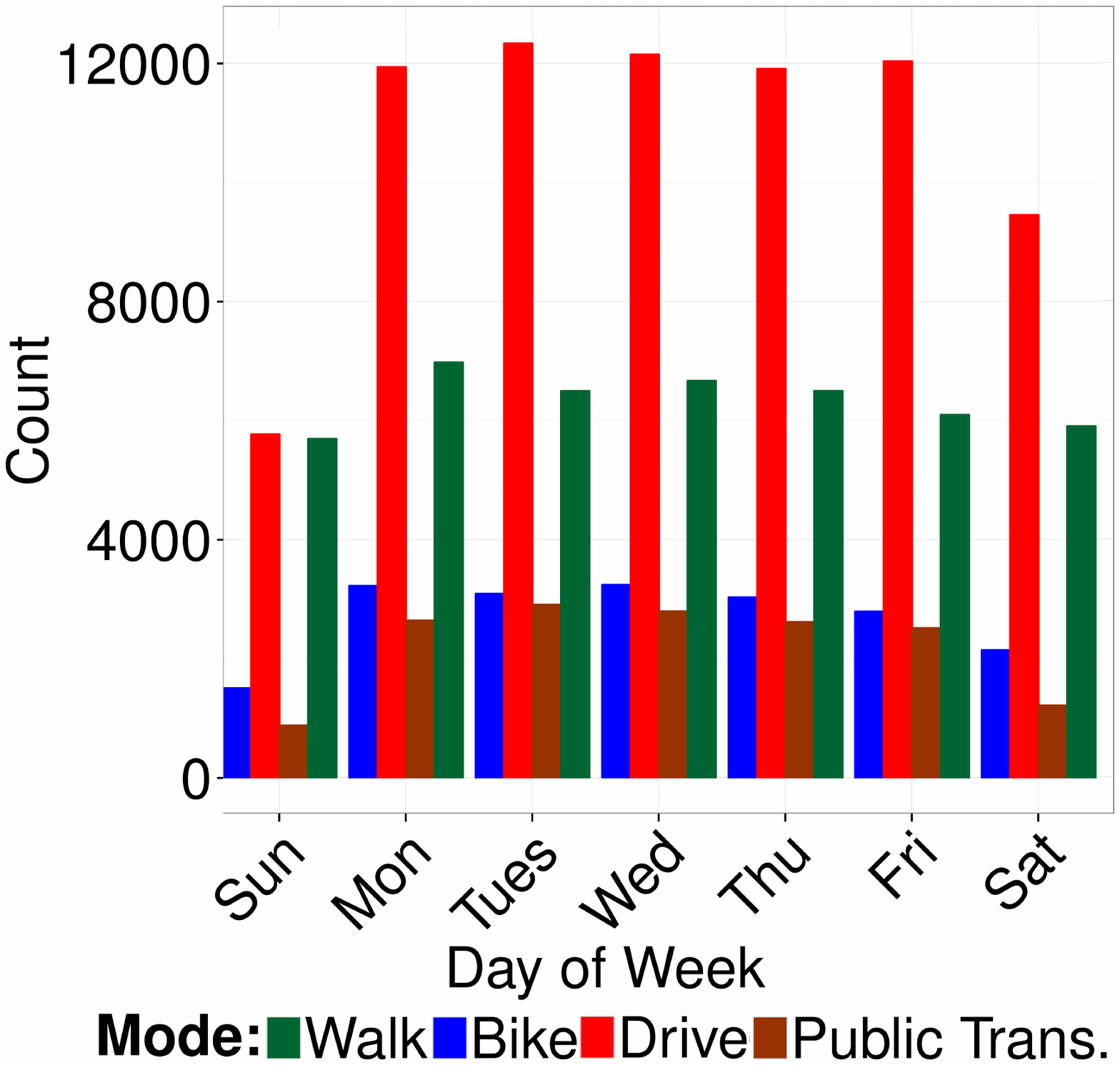} %./_st_stdg_trip_length.pdf}
      \label{fig:mode_share_day_of_week}
    }
    \subfloat[]{%
    \includegraphics[trim =9mm 55mm 9mm 45mm, clip, width=0.24\textwidth]{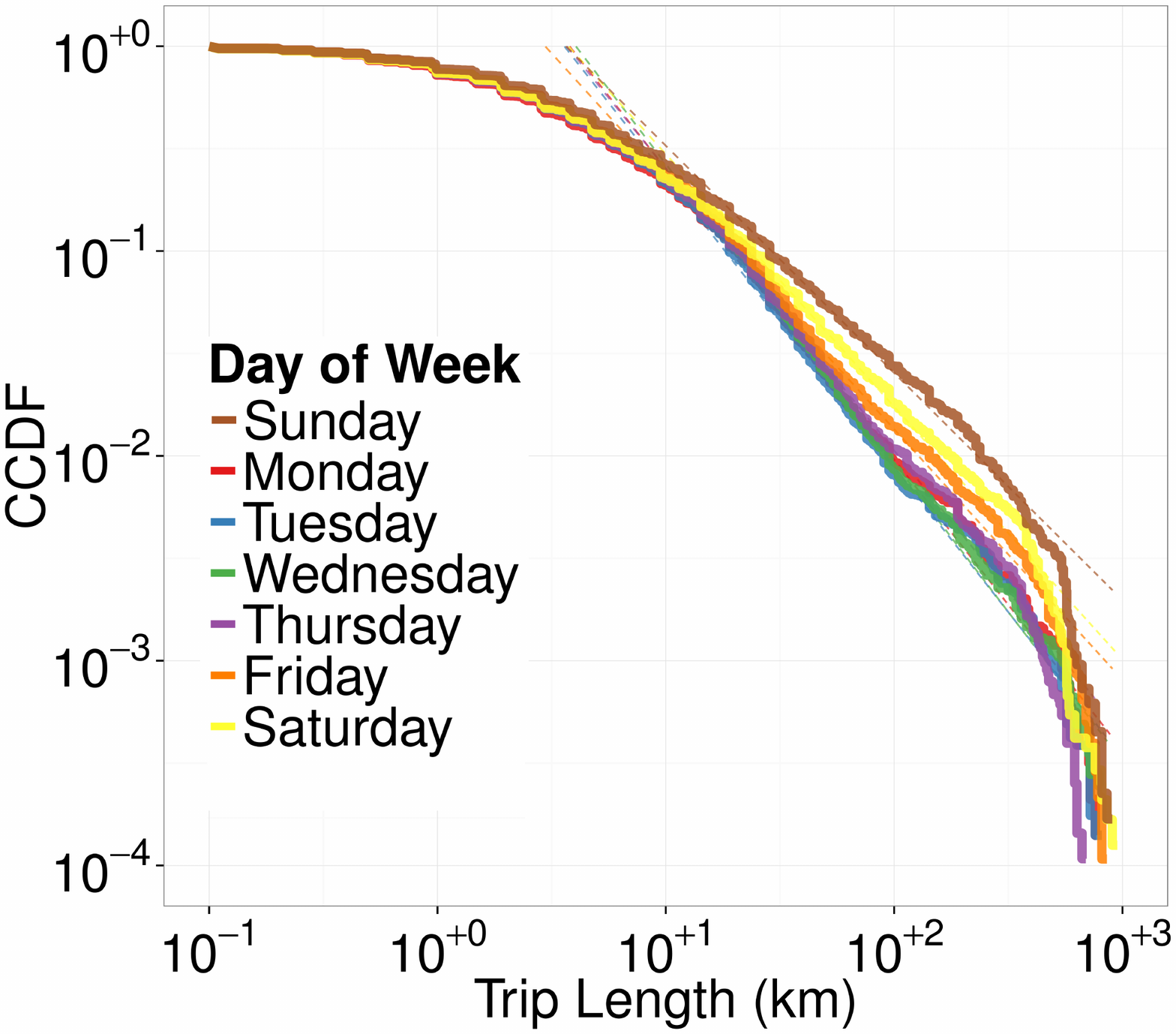} %./_st_stdg_trip_length.pdf}
      \label{fig:day_of_week_trip_lengths}
    }
     \caption{(\ref{fig:mode_share_hourly})  Weekday hourly trip frequency according to mode.  (\ref{fig:hourly_trip_lengths}) CCDF of weekday hourly trip lengths.  (\ref{fig:mode_share_day_of_week}) Day-of-week trip frequency according to mode. (\ref{fig:day_of_week_trip_lengths}) CCDF of trip lengths by day-of-week.}
     \label{fig:time_dependence}
    \end{figure}
%\vspace*{-.1cm}

\begin{table}[!ht]%\footnotesize
   % \vspace*{-.1cm}
\centering
\begin{tabular}{|l|c|c|c|c|c|c|}
\hline
Time of Day & Count & $\alpha$ & $\ell_0$ (km) & $\bar{\ell}$ (km) & $\sigma^2$\\
\hline
%cat.val & num.trips & alpha & x.min & av & v1
%cat.val & num.trips & alpha & x.min & av & v1
before 5 AM & 1670 & 2.01 & 25.27 & 32.92 & 10,123.99\\
\hline
5 to 7 AM & 7026 & 2.41 & 20.58 & 23.71 & 3,268.39\\
\hline
7 to 9 AM & 21991 & 2.33 & 16.15 & 11.29 & 1,717.64\\
\hline
9 to 11 AM & 24511 & 2.03 & 10.45 & 11.31 & 2,159.07\\
\hline
11 to 2 PM & 37693 & 2.32 & 31.36 & 9.49 & 1,046.26\\
\hline
2 to 5 PM & 43375 & 2.43 & 51.30 & 10.34 & 868.00\\
\hline
5 to 8 PM & 34742 & 2.55 & 31.36 & 9.68 & 705.90\\
\hline
8 to 10 PM & 7819 & 2.39 & 34.30 & 9.58 & 684.20\\
\hline
after 10 PM & 4060 & 2.89 & 30.40 & 10.94 & 550.62\\
\hline
\end{tabular}
\caption{Count, $\alpha$, $\ell_0$, $\bar{\ell}$ and $\sigma^2$ for trips by time of day.}
\label{table:hourly_fits}
\end{table}
%\vspace*{-.1cm}

\begin{table}[!ht]%\footnotesize
    \vspace*{-.2cm}
\centering
\begin{tabular}{|l|c|c|c|c|c|c|}
\hline
Day of Week & Count & $\alpha$ & $\ell_0$ (km) & $\bar{\ell}$ (km) & $\sigma^2$\\
\hline
%cat.val & num.trips & alpha & x.min & av & v1
Sunday & 17768 & 2.11 & 32.34 & 15.84 & 2,652.07\\
\hline
Monday & 28476 & 2.42 & 34.20 & 9.66 & 1,026.79\\
\hline
Tuesday & 28449 & 2.42 & 38.81 & 9.47 & 919.62\\
\hline
Wednesday & 28649 & 2.46 & 48.45 & 9.86 & 966.94\\
\hline
Thursday & 27787 & 2.38 & 38.95 & 10.07 & 943.46\\
\hline
Friday & 27878 & 2.22 & 43.23 & 11.46 & 1,507.96\\
\hline
Saturday & 23880 & 2.23 & 32.30 & 12.60 & 1,789.28\\
\hline
\end{tabular}
\caption{Count, $\alpha$, $\ell_0$, $\bar{\ell}$ and $\sigma^2$ for trips by day of week.}
\label{table:daily_fits}
\end{table}
\vspace*{-.1cm}

Finally, we see that trip frequency, mode share, and trip lengths are clearly dependent on time.  On weekdays, according to time-of-day, we see an expected daily pattern of increased trips in morning $(\sim7$ AM) and evening $(\sim5$ PM) (Fig. \ref{fig:mode_share_hourly}).  We also note a change in the relative mode share at different times of day.  Driving, for example, makes up a much higher proportion of trips during the day, lower in evening hours.  Trip lengths are also notably responsive to time-of-day, e.g. from 5 to 7AM, trips tend to be longer (Fig. \ref{fig:hourly_trip_lengths}).

Similar observations can be made about day-of-week patterns.  For example, on Sundays we see a change in trip frequency and mode share from weekday levels, with fewer overall trips and less driving relative to other modes.  (Fig. \ref{fig:mode_share_day_of_week}).  Trip lengths are also clearly responsive to day-of-week, with a higher proportion of long trips also on Sunday (Fig. \ref{fig:day_of_week_trip_lengths}).

Aggregation over all time periods can therefore also obscure time-dependency and potentially bias results.  We must conclude that sampling time needs thorough investigation when making statements characterizing average mobility patterns.
			
%     \begin{figure}[!ht]
%
%    \subfloat[]{
%    %%trim option's parameter order: left bottom right top
%    \includegraphics[trim =0mm 55mm 0mm 0mm, clip, width=0.35\textwidth]{./checkins_II.pdf}
%	\label{fig:checkins}}
%     \caption{\ref{fig:checkins} The actual path from A to B (black), conventional trips lengths reported (red, {$L_{C1}$,$L_{C2}$,$L_{C3}$}), and big mobility length reported ($L_B$)}
%    \end{figure}

%FIX LANGUAGE, LEGEND; REPLACE BOTTOM Fig. WITH DAY-OF-WEEK PLOT.  FIX BAR WIDTHS
%
%Note the change in overall frequency and mode share over time.
%
%Trip length distributions are notably responsive to time of day.
%
%WE HAVE ACTUAL TRIP LENGTHS, THEY HAVE LOWER BOUND ON TRIP LENGTHS, BUT NOT ACTUAL TRIP LENGTHS
%
%%We note that, if we treat both big mobility and conventionally-collected mobility data as power-law distributions, the ratio between them is also a power law,
%%
%%$$
%%p(r) = \frac{C r^{-\alpha}}{D r^{-\beta}} = \frac{C}{D}r^{\beta -- \alpha}.
%%$$
%
%
%We also note that the sampling rate can affect the apparent length of trips, and thereby the trip length distribution.

%As our data was collected explicitly to know trip lengths, we believe this explains the steeper slope of our trip lengths, so that $L_2$ will be divided into several trips in conventional data.

\section{Conclusions and Further work}

We have argued that aggregate data misses important aspects of mobility patterns. As a case study, we have analyzed a category-rich set of German mobility data and found that mode, city size, population, purpose, and temporal aspects of trips can be illustrative.  This conventional data can expose both inter- and intra- urban-scale mobility, and possibly address related issues such as urban metabolism, allometric scaling, and the debate between distance- and intervening-opportunity-based mechanisms for mobility patterns.

We understand our work as a first step toward a more refined understanding.  In particular, we have only focused on Germany and will be interested whether other countries have similar characteristics.  Our data may still have some bias and errors, and we would like to address those.  Moreover, so far we have focused on data analysis only.  In future work, it would be interesting to come up with models explaining the observed statistics.  Based on our work, mode, purpose, urban population, and time look like useful categories to investigate.  From other research, density, mode availability, and other urban parameters also seem relevant \cite{Noulas2011, Scheiner2010, Batty2010}.  Further work fitting trip length along with duration, analyzing mean squared distance, and using clustering and dimensionality reduction to understand the main categories and dependencies making up the space of mobility universality classes all seem promising.

%\subsection{Walk, Bike, Rideshare by Category}
%
%
%
%
%\begin{figure}[!ht]
%\includegraphics[scale=.3]{./walk_hwzweck_trip_length.pdf}
%\includegraphics[scale=.3]{./bike_hwzweck_trip_length.pdf}
%\caption{purpose is very important to walking and bicycling}
%\label{fig:walk_bike_drive_purpose}
%\end{figure}
%
%
%plots of walking, biking by zweck
%
%
%
%We also observe that scaling of modes has a implication for scaling of large energy consumption.  The slowly declining heavy tail of long automotive trips means that the `tail' of these trips plays a significant role in energy consumption.  Therefore the scaling of these trips is very significant in a systemic context for an understanding of urban energy.
%
%We also note that mode choice implies a different energy consumption budget.  Trivially, walking, and automobile modes have significantly different energy characteristics.  This has large economic and climate change implications, when we consider the universality of these patterns over countries or continents.

%\section{Acknowledgments}
%
%
%Thanks to Federico Camboni, Nathan Eagle, Marta Gonzalez, Pan Hui, Hamed Ketabdar, Hartmut Lentz, Arne Ludwig, Gregoire Montavon, Zoltan Neda, Felix Poloczek, Andreas Sorge, Shahin Tajik, Max Thess, Marc Timme and Peter Wagner for discussions and feedback.\\

\bibliographystyle{IEEEtran}

\bibliography{refs_short}

\begin{thebibliography}{10}
\providecommand{\url}[1]{#1}
\csname url@samestyle\endcsname
\providecommand{\newblock}{\relax}
\providecommand{\bibinfo}[2]{#2}
\providecommand{\BIBentrySTDinterwordspacing}{\spaceskip=0pt\relax}
\providecommand{\BIBentryALTinterwordstretchfactor}{4}
\providecommand{\BIBentryALTinterwordspacing}{\spaceskip=\fontdimen2\font plus
\BIBentryALTinterwordstretchfactor\fontdimen3\font minus
  \fontdimen4\font\relax}
\providecommand{\BIBforeignlanguage}[2]{{%
\expandafter\ifx\csname l@#1\endcsname\relax
\typeout{** WARNING: IEEEtran.bst: No hyphenation pattern has been}%
\typeout{** loaded for the language `#1'. Using the pattern for}%
\typeout{** the default language instead.}%
\else
\language=\csname l@#1\endcsname
\fi
#2}}
\providecommand{\BIBdecl}{\relax}
\BIBdecl

\bibitem{sousaaura}
J.~Sousa and D.~Garlan, ``Aura: An architectural framework for user mobility in
  ubiquitous computing environments,'' in \emph{Software Architecture: System
  Design, Development, and Maintenance (Proceedings of the 3rd Working
  IEEE/IFIP Conference on Software Architecture)}, 2002.

\bibitem{Vespignani2010}
A.~Vespignani, ``{Complex networks: The fragility of interdependency},''
  \emph{Nature}, vol. 464, 2010.

\bibitem{Griepp2013}
T.~Camp, J.~Boleng, and V.~Davies, ``A survey of mobility models for ad hoc
  network research,'' \emph{Wireless communications and mobile computing},
  vol.~2, no.~5, 2002.

\bibitem{Cho:2011:FMU:2020408.2020579}
E.~Cho, S.~A. Myers, and J.~Leskovec, ``{Friendship and mobility: user movement
  in location-based social networks},'' in \emph{Proceedings of the 17th ACM
  SIGKDD international conference on Knowledge discovery and data mining
  (KDD)}, 2011.

\bibitem{Silva2006}
B.~C. da~Silva, A.~L. Bazzan, G.~K. Andriotti, F.~Lopes, and D.~de~Oliveira,
  ``Itsumo: an intelligent transportation system for urban mobility,'' in
  \emph{Innovative Internet Community Systems}.\hskip 1em plus 0.5em minus
  0.4em\relax Springer, 2006.

\bibitem{Helbing2004}
D.~Helbing and K.~Nagel, ``{The physics of traffic and regional development},''
  \emph{Contemporary Physics}, vol.~45, no.~5, 2004.

\bibitem{Chen2006}
X.~Chen and F.~B. Zhan, ``Agent-based modelling and simulation of urban
  evacuation: relative effectiveness of simultaneous and staged evacuation
  strategies,'' \emph{Journal of the Operational Research Society}, vol.~59,
  no.~1, 2006.

\bibitem{Helbing2000}
D.~Helbing, I.~Farkas, and T.~Vicsek, ``Simulating dynamical features of escape
  panic,'' \emph{Nature}, vol. 407, no. 6803, 2000.

\bibitem{Momoh2009}
J.~A. Momoh, ``Smart grid design for efficient and flexible power networks
  operation and control,'' in \emph{IEEE/PES Power Systems Conference and
  Exposition (PSCE), 2009.}

\bibitem{Townsend_2000}
A.~Townsend, ``{Life in the real-time city: Mobile telephones and urban
  metabolism},'' \emph{Journal of urban technology}, vol.~7, no. 212, 2000.

\bibitem{banister2011cities}
D.~Banister, ``Cities, mobility and climate change,'' \emph{Journal of
  Transport Geography}, vol.~19, no.~6, pp. 1538--1546, 2011.

\bibitem{Brockmann:2006uq}
D.~Brockmann, L.~Hufnagel, and T.~Geisel, ``{The scaling laws of human
  travel},'' \emph{Nature}, vol. 439, no. 7075, 2006.

\bibitem{Gonzalez:2008uq}
M.~C. Gonzalez, C.~A. Hidalgo, and A.-L. Barabasi, ``{Understanding individual
  human mobility patterns},'' \emph{Nature}, vol. 453, no. 7196, 2008.

\bibitem{Noulas2011}
A.~Noulas, S.~Scellato, R.~Lambiotte, M.~Pontil, and C.~Mascolo, ``{A tale of
  many cities: universal patterns in human urban mobility},'' \emph{PloS one},
  vol.~7, no.~5, 2012.

\bibitem{simini_2012}
F.~Simini, M.~Gonz\'{a}lez, A.~Maritan, and A.~Barab\'{a}si, ``{A universal
  model for mobility and migration patterns},'' \emph{Nature}, vol. 484,
  no.~96, 2012.

\bibitem{Stouffer1940}
S.~Stouffer, ``{Intervening opportunities: a theory relating mobility and
  distance},'' \emph{American sociological review}, 1940.

\bibitem{song2010modelling}
C.~Song, T.~Koren, P.~Wang, and A.-L. Barab{\'a}si, ``Modelling the scaling
  properties of human mobility,'' \emph{Nature Physics}, vol.~6, no.~10, 2010.

\bibitem{Scheiner2010}
J.~Scheiner, ``{Interrelations between travel mode choice and trip distance:
  trends in Germany 1976-2002},'' \emph{Journal of Transport Geography},
  vol.~18, no.~1, 2010.

\bibitem{Newman2009}
A.~Clauset, C.~Shalizi, and M.~Newman, ``{Power-law distributions in empirical
  data},'' \emph{SIAM review}, 2009.

\bibitem{Song2010}
C.~Song, Z.~Qu, N.~Blumm, and A.-L. Barab\'{a}si, ``{Limits of predictability
  in human mobility.}'' \emph{Science (New York, N.Y.)}, vol. 327, no. 5968,
  2010.

\bibitem{Reddy2010}
S.~Reddy, M.~Mun, J.~Burke, D.~Estrin, M.~Hansen, and M.~Srivastava, ``Using
  mobile phones to determine transportation modes,'' \emph{ACM Transactions on
  Sensor Networks (TOSN)}, vol.~6, no.~2, 2010.

\bibitem{Follmer2008a}
R.~Follmer, D.~Gruschwitz, B.~Jesske, and S.~Quandt, ``{Mobilit\"{a}t in
  Deutschland 2008},'' Bundesministerium fuer Verkehr, Bau, und
  Stadtentwicklung, Bonn, Tech. Rep., 2008.

\bibitem{newman2005power}
M.~E. Newman, ``Power laws, pareto distributions and zipf's law,''
  \emph{Contemporary physics}, vol.~46, no.~5, 2005.

\bibitem{wolman1965metabolism}
A.~Wolman, ``The metabolism of cities,'' \emph{Scientific American}, vol. 213,
  1965.

\bibitem{west2004life}
G.~B. West and J.~H. Brown, ``Life's universal scaling laws,'' \emph{Physics
  today}, vol.~57, no.~9, 2004.

\bibitem{Office}
\BIBentryALTinterwordspacing
``{St\"{a}dte (Alle Gemeinden mit Stadtrecht) nach Fl\"{a}che, Bev\"{o}lkerung
  und Bev\"{o}lkerungsdichte},'' German Federal Statistical Office, Tech. Rep.,
  2012. [Online]. Available:
  \url{https://www.destatis.de/DE/ZahlenFakten/LaenderRegionen/Regionales/Geme%
indeverzeichnis/Administrativ/Aktuell/05Staedte.html}
\BIBentrySTDinterwordspacing

\bibitem{Mitzenmacher2004}
M.~Mitzenmacher, ``{A Brief History of Generative Models for Power Law and
  Lognormal Distributions},'' \emph{Internet Mathematics}, vol.~1, no.~2, 2004.

\bibitem{Stanley1999}
H.~Stanley, ``{Scaling, universality, and renormalization: Three pillars of
  modern critical phenomena},'' \emph{Reviews of modern physics}, vol.~71,
  no.~2, 1999.

\bibitem{ramos2004levy}
G.~Ramos-Fernandez, J.~L. Mateos, O.~Miramontes, G.~Cocho, H.~Larralde, and
  B.~Ayala-Orozco, ``L{\'e}vy walk patterns in the foraging movements of spider
  monkeys (ateles geoffroyi),'' \emph{Behavioral Ecology and Sociobiology},
  vol.~55, no.~3, 2004.

\bibitem{Helbing2007}
L.~M. Bettencourt, J.~Lobo, D.~Helbing, C.~K{\"u}hnert, and G.~B. West,
  ``Growth, innovation, scaling, and the pace of life in cities,''
  \emph{Proceedings of the National Academy of Sciences}, vol. 104, no.~17,
  2007.

\bibitem{adac_reisemonitor_2008}
``{Reisemonitor 2008},'' ADAC Verlag, Tech. Rep., 2008.

\bibitem{Batty2010}
M.~Batty, ``{Spatial Entropy},'' \emph{Geographical Analysis}, vol.~6, no.~1,
  2010.

\end{thebibliography}

\end{document}